# Being good (at driving): Characterizing behavioral expectations on automated and human driven vehicles

*Laura Fraade-Blanar, Francesca Favarò, Johan Engstrom,
Melissa Cefkin, Ryan Best, John Lee, Trent Victor*

**Abstract:**[1]
The question of how to define good driving is not new. For over a century, researchers have wrestled with an answer for human drivers, and the debate has recently surfaced afresh for automated vehicles (AVs). Although numerous principles and models have striven to explore the associated complexities, no framing exists to coordinate and align these concepts into a clear vision.

In response to these challenges, we put forth the concept of Drivership as a framing for the realization of good driving behaviors. Drivership grounds the evaluation of driving behaviors in the alignment between the mutualistic expectations that exist amongst road users. Leveraging existing literature, we start by distinguishing (i) Empirical Expectations (i.e., reflecting "beliefs that a certain behavior will be followed," drawing on past experiences) (Bicchieri, 2006); and (ii) Normative Expectations (i.e., reflecting "beliefs that a certain behavior ought to be followed," based on societally agreed-upon principles) (Bicchieri, 2006). Because societal expectations naturally shift over time, we introduce a third type of expectation, Furtherance Expectations, denoting behavior which could be exhibited to enable continuous improvement of the transportation ecosystem. We position Drivership within the space of societal Normative Expectations, noting an existing overlap with some Empirical and Furtherance Expectations, constrained by what is technologically and physically feasible.

More generally, we establish a novel vocabulary to more rigorously tackle the conversation on stakeholders' expectations, which is a key feature of value-sensitive design approaches. We also detail how Drivership comprises safety-centric behaviors, but extends beyond those, to include what we here term socially-aware behaviors (where there are no clear safety stakes).

Drivership supports multiple purposes, including advancing the understanding and evaluation of driving performance through benchmarking based on many criteria. As such, we argue that an appropriate framing of the notion of Drivership also underpins the overall development of a safety case. The paper explores these applications under the more general tenet of Drivership as a central element to roadway citizenship.

## 1. Introduction

On September 13, 1899 in New York City near Central Park, Mr. Henry Hale Bliss alighted from a streetcar and was assisting his companion to exit when he was hit by a car. Because of a truck parked to the right of the lane, the car was driving unusually close to the streetcar. Mr. Bliss died the following day, making him the first U.S. automotive crash fatality. The car's driver was charged with manslaughter and later acquitted as the crash was deemed "unintentional." (Williams, 2018; Eschner, 2017).

---

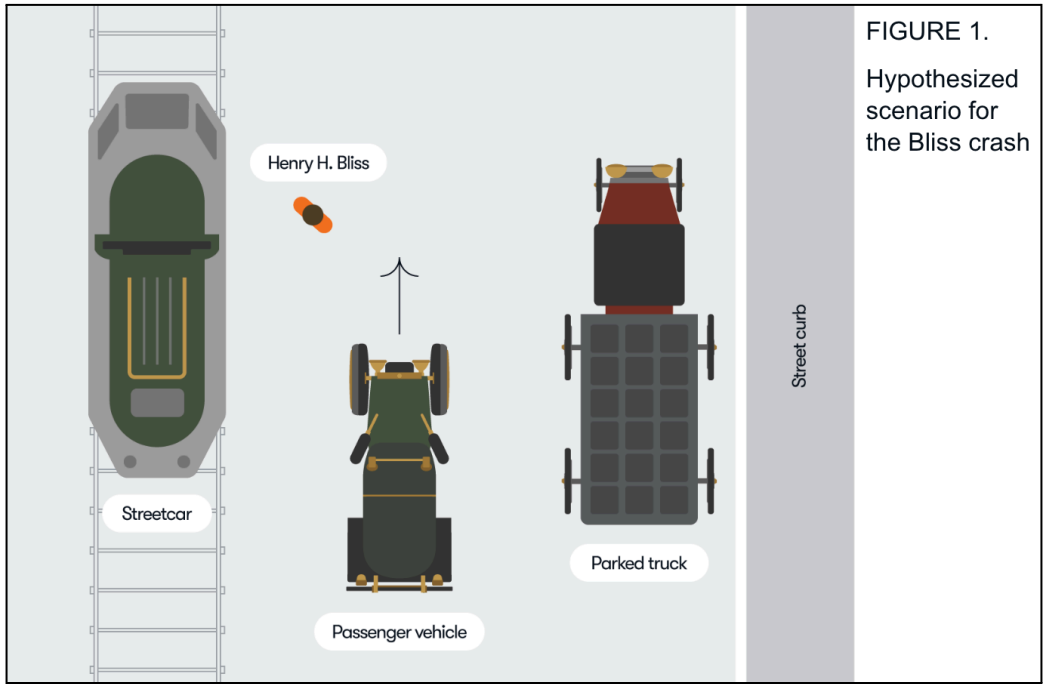

FIGURE 1. Hypothesized scenario for the Bliss crash

Consider the scene for the 1899 Bliss crash (Figure 1).[2] The driver's response to a potential conflict (and, possibly, a late reveal) resulted in a negative outcome; yet, depending on the vehicle mechanics, a better response may not have been possible. No traffic laws were broken: New York City did not enact their first traffic code until 1903 (American Safety Council, n.d.). Speeding enforcement was already underway (starting in 1895) with a primary focus on bicycles and a limit of 8 mph, (NYPD, n.d.),[3] but there is no record of the car driving inappropriately fast or recklessly. It could be argued that the driver should have anticipated the risk of passengers stepping out from the streetcar and, hence, should have slowed upon entering the narrow passage in the first place (assuming he did not). The truck's parking choice may have been inconsiderate, given it impacted traffic flow. Mr. Bliss may have unintentionally initiated the conflict by stepping out into oncoming traffic (as a consequence of the streetcar stopping where it did[4]), but he stepped out onto a set streetcar route where road users perhaps should expect exiting pedestrians. Although 125 years have passed since Mr. Bliss's death, this type of scenario and the questions it generates still resonate today. How can the "goodness" of each agent's behavior be defined and evaluated? How can this scenario be understood and used to learn and improve behavior within the transportation ecosystem? And why is this important?

---

[2] Visuals of the crash scenario are not available, so this scene has been reimagined from existing narratives and photographs of period-specific streetcar pick-usp and drop-offs.

[3] New York City's first speeding law, enacted in 1652, limited horses to a speed less than a gallop (American Safety Council, n.d.).

[4] Although this paper focuses on road user behavior, it is important to note the role that infrastructure plays in influencing behavior. According to contemporary news accounts, "the place where the accident happened is known to the motormen on the streetcar line as 'dangerous stretch,' on account of the many accidents which have occurred during the past summer." ("Fatally Hurt by Automobile," 1899).

These open questions illustrate the challenge of understanding and assessing good driving behavior of autonomous/automated driving systems and human drivers.[5] To find answers, this paper starts from existing definitions of driving behavior, reviewing the literature to identify different qualifications of goodness and exploring the complexities associated with this determination. In response, we put forth the concept of Drivership,[6] the realization of good driving behavior, predicated on the alignment of driving behavior and societal expectations on driving. Although societal expectations may differ in application between human drivers and automated driving systems, Drivership is driver-agnostic (i.e., applying to both human and non-human drivers). We argue that, within Drivership, identifying and meeting (or exceeding) these expectations entails an analysis that includes safe behavior but also behavior where there are no safety stakes. Furthering this, we disaggregate societal expectations into subtypes through a taxonomic approach, to support further discussion. Finally, we conclude by diving into the question of how Drivership relates to a safety case and a top goal of absence of unreasonable risk.

## 2. What is good driving behavior?

Considering the purpose of good driving behavior necessitates an ethics-based exploration (Gerdes, 2020). Amidst the ethical landscape, a resonating perspective[7] positions good driving behavior as the realization of balancing a communal expectation for road-users to respect each other and be similarly respected when sharing a common road, given consensus on the utility and role of transportation in current society. For example, what responsibility does an ADS developer or fleet operator have to their customers, from one-time users to those who depend on ADSs for their independence and spontaneity, and what is owed to others with whom they share the roadway? What do other road users owe the ADS, given that driving involves public actions occurring in shared spaces? How do disparate goals align to give rise to moral principles on driving behavior which "no one (could) reasonably reject?" (Scanlon, 2010). What "mutual recognition" (Ashford & Mulgan, 2007) and mutual constraints should be encouraged or required between road users to provide for a positive shared road experience?

Good driving behavior contributes to roadway citizenship. Roads and adjacent spaces are often public spaces, governed by a specific set of rules and broader societal expectations[8] which, by entering into said space, users agree to abide by. A mutual vulnerability to each other's behavior on the roadway necessitates this mutual adherence and respect, beyond the general understanding of the applicable rules of the road.

Philosophically-inclined readers will recognize, here and throughout this text, tenets of T.M. Scanlon's philosophy of Contractualism (Scanlon, 1998). There is a long history of basing principles of good driving on philosophical tenets: e.g., d'Amato et al. (2022) refer to a Locke-ian (Uzgalis, 2001) social contract between road users and with road rules, and it could be argued that Responsibility-Sensitive Safety (RSS) (Shalev-Shwartz et al., 2018) is an example of deontological ethics (Geisslinger et al., 2021; Gerdes, 2020). Arguments for a positive risk balance are generally understood to be referencing utilitarianism (Geisslinger et al., 2021). More broadly, it has been argued that general ADS efforts to improve driving behavior are a form of Aristotelian virtue ethics (Geisslinger et al., 2021; Gerdes, 2020).

---

[5] Of note, this paper focuses on driving behavior rather than driving outcomes; leading rather than lagging (Fraade-Blanar et al., 2018). Crashes and deaths are the expression of road users' behavior in conjunction with a range of other factors, such as infrastructure, geography, emergency medical service proximity, etc.

[6] Inspired by the concept of Roadmanship (Fraade-Blanar et al., 2018), Drivership was first presented in 2022 at the Automated Road Transportation Symposium and at Lifesavers, with additional social media (Waymo, 2022).

[7] This refers to a perspective which may meet with common agreement.

[8] Societal expectations are discussed in detail in section 3.1.

**2.1 How has "good driving behavior" been defined, to date?**
Setting aside for the moment the qualifier "good," let's begin with "driving behavior." The technical literature on the sciences of driving and traffic performance[9] is vast. SAE J3164[10] defines driving behaviors as "goal-oriented activities enacted by a vehicle to implement a human driver's or an engaged ADS's (automated driving system) performance of the Dynamic Driving Task (DDT) or DDT fallback." Within current SAE documentation J3016 and J3164 (SAE, 2021; SAE, 2023), behavior itself has been leveled into (Michon, 1985):
- *Operational*: pertaining to the moment-to-moment vehicle control associated with maintaining speed, lateral lane position, and separation between vehicles,
- *Tactical:* pertaining to the ability to make decisions about if and when one should optimally overtake another vehicle, change lanes, select a target speed, etc.,
- *Strategic:* pertaining to broader considerations such as the choice of route and the decision to make a trip.

Several taxonomies have striven to dissect driving behaviors into specific actions or tasks, at times also encompassing safety relevancy and ranking across such tasks. McKnight et al. (1972) identified and then ranked driving actions by criticality, noting that some driver actions are more safety critical than others. The goal was to improve education and generate safer drivers, an effort repeated over the years to similar results (e.g., Bouhsissin, Sael, and Benabbou, 2023).

Let us now turn our focus to what makes driving behaviors "good." The recognition that certain elements of driving behavior are essential for safety represents one of the key ideas behind the concept of *Roadmanship*, introduced in 2018. Defined as a leading measure of safety, it describes "the ability to drive on the road safely without creating hazards and responding well… to the hazards created by others. The concept centers on whether the vehicle 'plays well with others,' even if others are not around," (Fraade-Blanar et al., 2018). The original text included a range of safety-specific metrics such as time-to-collision, maintaining a safety envelope (as used in IEEE 2846 (IEEE, 2022) and defined in ADSSC's Best Practice for Metrics and Methods for Assessing Safety Performance of Automated Driving Systems (ADS) (AVSC, 2021)), and hard braking, each metric with their own strengths and weaknesses and each drawing from decades of safety research (Fraade-Blanar et al., 2018). Roadmanship also highlighted the importance of conflict avoidance (defensive driving) and secondary crashes[11] (Fraade-Blanar et al., 2018).

---

[9]This report does not aim to be comprehensive of all related literature, but rather notes key texts.

[10] This SAE Information Report is titled J3164 Ontology and Lexicon for Automated Driving System (ADS)-Operated Vehicle Behaviors and Maneuvers in Routine/Normal Operating Scenarios (SAE, 2023).

[11]A secondary crash is a crash that occurs as a result of an original crash either within the crash scene or within the queue or backup in either direction." (Pecheux et al., 2023)

The work on Roadmanship refocused a general interest in behavioral safety for the ADS community and ignited new research on the operationalization of technical definitions for the behaviors and maneuvers[12] which directly reflect "good" driving. At the same time, it made clear that good driving behavior included more than just safe driving behavior. To date, much work connecting "good" to specific behaviors or considering behavioral goals focuses on safety or legal compliance with rules of the road. The latter refers to the local and state laws and regulations that determine how drivers and other road users should behave on the road – *what* road users should do, *how* they should do it, and *where* they are when they do it.

There is considerable work on road rules from industry and academia. D'Amato et al. (2022) focused on the aforementioned "social contract for driving," supported by traffic code, driving and case law, etc. " Similarly, Xiao et al., (2021) focused on identifying desirable driving from a legal viewpoint. Together these and other works suggest how the traffic code both influences and is influenced by social and ethical expectations. Specifically: "to consider the road rules of the future we start with the rules of the present. Behaviour on the road can be seen as governed by rules in four layers: (1) the laws of physics can be seen as constraints on what you can and cannot do; (2) legal rules define what you must and must not do; (3) advisory guidance shapes what you should and should not do; and (4) then there is a layer of social norms and behaviours: what we tend to do and not do. These layers overlap. Many of our social norms align with the rules of the other layers." (Tennant et al., 2021)

Rooted in the second layer, the California Department of Motor Vehicle's regulations for the Deployment of Autonomous Vehicles stipulate the following:
> *(9) The manufacturer shall certify that the autonomous technology is designed to detect and respond to roadway situations in compliance with all provisions of the California Vehicle Code and local regulation applicable to the performance of the dynamic driving task in the vehicle's operational design domain, except when necessary to enhance the safety of the vehicle's occupants and/or other road users.* (CA DMV, 2022)

Importantly, this type of regulatory provision recognizes the importance of automated vehicles satisfying the expectations of road rule compliance, while also prioritizing actions for enhancing safety if conflicts between competing road rules (Xiao et al., 2021), or road rules and safety, emerge for that operational design domain.

Rooted in the third and fourth layer, the approach in ISO 39003 Ethical Considerations Relating to Safety for Autonomous Vehicles is based not on concepts (e.g., safety or compliance with rules of the road) but rather on specific principles:[13] some situational (e.g., Driving Rule DR6: *Make lane changes only if there is an empty 'slot' and there are no dangerous … consequences,* and DR3: *Even if there are no other vehicles in the vicinity, the vehicle should always be prepared for the sudden appearance of other vehicles with respect to its actions*.), some value-based (e.g., Driving Rule DR4: *During every ride or sortie, show respect and kindness to other road users*, and DR8-a: *Act in a way that acknowledges, communicates with,*

---

[12] Per SAE J3164, a vehicle maneuver is a "goal-oriented vehicle movement directed by an ADS or human driver in order to achieve a specific result/outcome." (SAE, 2023)

[13] One of the sources leveraged to build consensus in the ISO technical committee on road safety management (which produced ISO 39003) was the German Ethics Commission on Automated and Connected Driving's report (Fabio et al., 2017), which focuses on good behavior, noting the importance of preventing collisions and, as such, a prioritized approach to safety concerns.

*and respects every entity in the road space.*), etc. (ISO, 2023). Unfortunately, taken together, these principles, while laudable, provide little guidance for practical application.

Rooted in the second, third, and fourth layers, the academic literature contains multiple qualifications for good driving behavior, some of which conflict, many of which align, and all of which incentivize slightly-different to very-different behaviors. Among the clearest is the SPRUCE (safe, predictable, reasonable, uniform, comfortable, and explainable) model (De Freitas et al., 2021), and Vinkhuyzen and Cefkin's (2016) social acceptability-focused criteria, describing driving that would "smoothly integrate into the flow of traffic and handle roadway interactions without disrupting other road users." IEEE 7000,[14] in annex G, provides a wide range of ethical values such as care, fairness, politeness, respectfulness, etc. (IEEE, 2021) on which one could draw, although not all apply to driving behavior. Other studies used interviews with drivers to conclude that a good driver was characterized by "self-confidence, lack of stress, and not being aggressive," (Doubek et al., 2021), by comfort and naturalness (Peng et al., 2024), or by the interviewee's own personal driving behavior (Roy & Liersch, 2013). Reddy et. al. (2024) described a careful and competent driver as one who embodied "awareness, discernment, adaptability, patience, anticipation, caution, roadmanship, responsiveness, communication, and coordination." Still other proposed qualifications for goodness include attributes such as: empathetic (Nordfjærn & Şimşekoğlu, 2014), courteous and self-aware (Tillmann & Hobbs, 1949), just courteous (Wang et al., 2019), assertive (Oliveira, 2019), trustworthy (Lee & See, 2004), compliant (Bin-Nun et al., 2022), fuel-efficient (Wijayasekara et al., 2014), and adaptable to other road user behavior (Schwarting et al., 2019).

There have been some limited attempts in the literature to operationalize individual driving qualities to assess goodness.[15] For example, prioritizing smoothness of movement (Wang et al., 2018), indicating competence and confidence (Alsaid et al., 2020), can be assessed through measuring the minimization of jerk (Itkonen et al., 2017). Or, to prioritize efficiency (Lee & Son, 2011), one could assess progress along the route compared to a daily average, reflected in speed choice and how quickly the vehicle negotiates an intersection. Or, a benchmark may focus on defensiveness (Peng et al., 2022), assessed (partially) by the rate of rapid acceleration or deceleration. Comfort could be measured by assessment of vehicle handling and specifically by roll angle (Uys et al., 2006), although evaluating vehicle handling remains a developing field (e.g., (Oh et al., 2021; Hang & Chen, 2021a)). In the face of this general fragmentation, no clear path arises. Specifically, pegging goodness to driving characteristics is problematic because of the lack of consensus on which characteristics are optimal, and because what is optimal may change based on circumstances and is dependent on context.

A different approach from the enumeration of contributory attributes is that of defining a formal reference for behaviors, known as a behavior reference model. In this approach, the behavior of the driver under evaluation (e.g., human driver or an ADS) is compared against a behavioral

---

[14] This is the IEEE Standard Model Process for Addressing Ethical Concerns during System Design, a standard for organizations generally (and is not specific to ADS behavior or to ADS developers) (IEEE, 2021).

[15] Guidance exists to support the process of operationalizing societal principles and necessary exceptions to specific, value-based actions (e.g., Townsend et al., 2022).

model representing some driving performance benchmark.[16,17] Efforts include, but are not limited to:

- **Safety envelope models** establish a formal behavioral model which, given certain assumptions, creates a theoretical guarantee (based on mathematical arguments) that no crashes would happen if the model is universally adopted by all drivers. Examples include Responsibility-Sensitive Safety (RSS) (Shalev-Shwartz et al., 2017), the Safety Force Field (Nister et al., 2019), models based on reachable sets (Althoff & Dolan, 2014; Pek et al., 2020) and/or other spatio-temporal constraints (Favaro et al., 2018);
- **Estimating the momentary risk of collision based on kinematic measures**, for example the Instantaneous Safety Metric (ISM) (Weng et al., 2020);
- **Operationalizing Roadmanship** around specific behaviors such as gap acceptance in roundabouts using naturalistic driving data as a benchmark (Peng et al., 2022);
- **Defining a human driver model** that represents a proper response to an impending hazard. Examples include the Careful and Competent Driver Model (CCDM) (Japan Automobile Manufacturers Association, Inc., 2022) which aims to represent the collision avoidance performance of a "good" (careful and competent) driver in select conflict scenarios, the Fuzzy Safety Model (FSM) (Mattas et al., 2020; 2022) which is intended to represent a driver that "can anticipate the risk of a collision and apply proportionate braking" (ECE, 2023), and the NIEON model (Engstrom et al., 2024) intended to represent the collision avoidance response performance of a human driver that is Non-Impaired and have their Eyes ON the conflict. The CCDM and FSM models have been included in UNECE Regulation 157 (R157) on Automated Lane Keeping Systems (ECE, 2023).

These efforts vary in usability. Existing implementations tend to be narrowly scoped (e.g., limited generalizability to diverse scenarios), with little to no overarching relationship to other reference models. Moreover, models vary in the quality of signal that they provide: the CCDM and FSM were evaluated by Olleja, Markkula and Bärgman (2024) with respect to their ability to predict human collision avoidance performance in the SHRP2 naturalistic driving dataset, testing the assumption that they properly represent a "careful and competent driver." For both models, it was found that the collision behavior they exhibited deviated significantly from that of human drivers, where CCDM generally responded to the hazard later than human drivers while the FSM responded earlier.

While a relatively large number of usable behavior reference models and other operationalizations of good driving behavior exists, an overarching framework is needed to develop consensus to guide operationalization and facilitate alignment on behavioral benchmarks (IEEE, 2024). Although the literature shows many ways to describe driving behavior, to describe goodness, and to measure driving behavior and compare it to a benchmark representing goodness, no framing exists to coordinate and align these concepts into a clear, cohesive vision (Reddy et al., 2024), in order to provide precise guidance on operationalization.

---

[16] A potential operationalization that does not use behavior reference models can be found in the digital commentary driving technique, wherein the ADS provides data continuously collected during a trip to create a driving narrative (British Standards Institution, 2021).

[17] These concepts are further explored in the in-progress SAE J3330 recommended practice on behavior reference models for evaluating automated driving systems in traffic conflict scenarios (SAE, 2024).

We propose a novel framing, grounding good driving behavior in realizing societal expectations. This framing is called Drivership. In this context, Drivership[18] becomes an expression of how mutualistic expectations between all road users are being met by drivers on the road.

## 3. What is Drivership?

The concept of Drivership[19,20] has been developed as a framing of the realization of good driving behavior. For ADSs, company-internal development and evaluation, shaped by an array of endogenous and exogenous factors, beget the behavior the ADS will exhibit on the road. Drivership research advances the ongoing process of aligning driving behavior and societal expectations,[21] engendering public trust and societal acceptance (Figure 2). Particular attention to alignment may be necessary as societal expectations and behavior change, e.g., in reaction to new technology.

---

[18] In this document, Drivership is primarily discussed from a passenger vehicle driver viewpoint, but that is not to say these are the only entity on which expectations are placed. There are societal expectations on the behaviors of pedestrians, bicyclists, truck drivers, etc., some of which are enacted into law.

[19] Drivership operates across all levels of driving behavior (Michon, 1985). Notably, at the tactical and operational level, driving behavior is constrained by the laws of physics; what is physically and technologically feasible.

[20] Drivership spans the pre-crash driving behaviors, and crash and post-crash behaviors (should a crash occur), drawing from the Haddon Matrix (US DOT, 2011). For example, maintaining proper motion control is essential to prevent a crash, and essential during and after a crash has occurred to minimize further harm.

[21] An important role of assessing Drivership is to surface conflicts between different social expectations and stakeholder expectations, and to guide resolution. Formalizing such tradeoffs could identify pathways to mitigation (Defense Science Board, 2012).

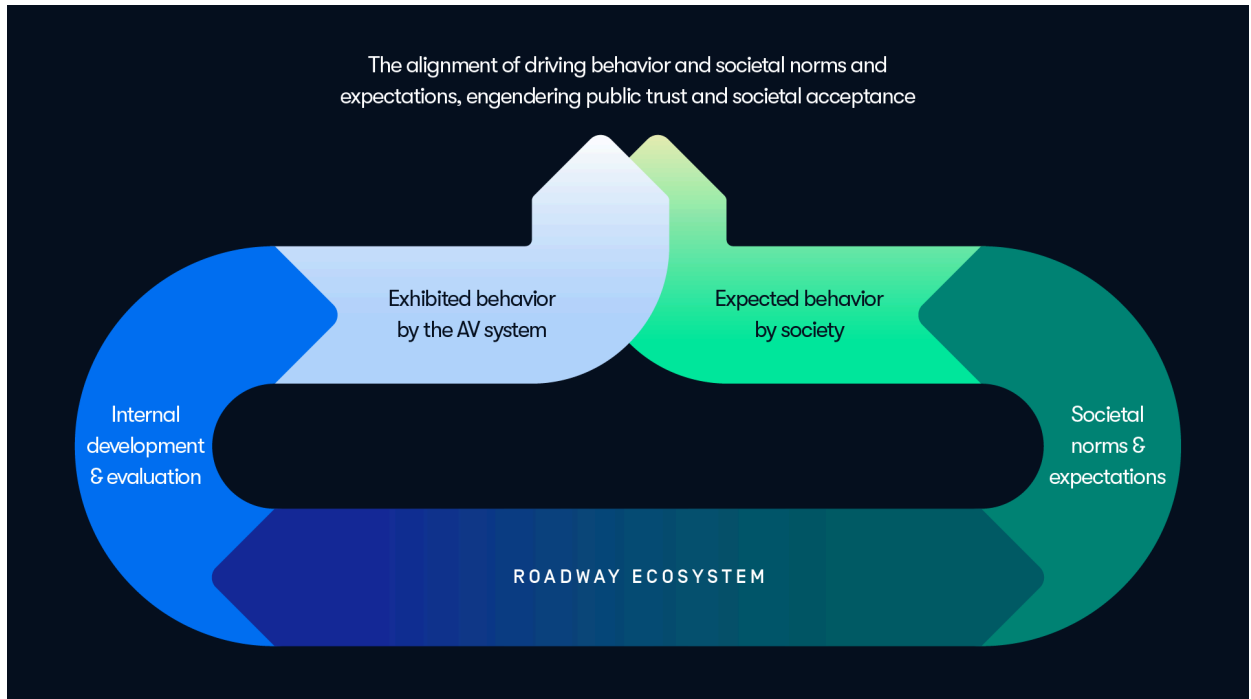

Figure 2. The alignment of driving behavior and societal norms and expectations for ADSs[22]

Safety lives at the core of Drivership, building on the aforementioned concept of Roadmanship (Fraade-Blanar et al., 2018). Roadmanship, which focuses only on safe driving behavior, and numerous other sources make clear that driving safely is a societal expectation for road users. Focusing on expectations of behavior that clearly and directly relate to safety is, however, *necessary but not sufficient* for the full realization of *"good"* driving behavior. Standards such as IEEE 7000 (2021) and publications described in section 2.1 above, make clear that expectations for the realization of good driving behavior extend beyond safety.

Consequently, we position Drivership as ranging from *safety-centric* to *socially-aware* behavioral expectations.[23] Safety-centric refers to expectations on behaviors which directly influence the probability of an injury. In a causal chain (as described by Favaro et al. (2023)), these are hazardous behaviors immediately proximal to harm. Conversely, socially-aware behaviors are more distal to harm manifestation, and without a clear or proximal injury risk. These behaviors may be more related to expectations around the social customs of driving. For example, a trajectory or proximity conflict (e.g., the subjective perception of cutting people off, or a pedestrian passing too closely) may be interpreted as uncourteous or normal depending on local social customs.[24] In describing driving behaviors, note that behaviors aligned with Drivership can have safety considerations but no social considerations, but not vice versa. In other words, Drivership cannot be realized through a courteous but unsafe behavior, but it remains possible to attain Drivership through a safe but uncourteous behavior.

---

[22] Social expectations may be deeply or lightly rooted in a given locality. The variation between localities and the likelihood of an expectation changing is an important topic for future discussion.

[23] Drivership draws from parallel concepts in other transportation modes, such as seamanship and airmanship. These contexts have a more lengthy and rich history of skilled vehicle operation and strictly apply to trained professionals. Additionally, the concepts of seamanship and airmanship are still somewhat nebulous and lack an agreed upon definition (e.g., (Kongsvik et al., 2020; Nergard, 2014). Though resistant to a precise definition, consistent themes of skilled operation in the context of the air and sea apply directly to the concept of Drivership. The base definition of both builds upon the precept that a thorough understanding of the rules of operation are a necessary but insufficient factor in establishing expertise. Airmanship and seamanship are communicated and demonstrated through action. Safe, reliable, and successful actions in routine trips serve to demonstrate understanding of the rules of operation, but similar actions in atypical scenarios communicate an important aspect of skilled operation – the ability to act in ways incongruent with rules, if and when necessary, to avoid negative outcomes.

The notion of Drivership is in line with what was proposed in the IEEE 7000 standard, which details processes by which organizations can include philosophical considerations in product development and refinement (IEEE, 2021). By taking into account societal expectations from trip start through trip end, there is opportunity to consciously and deliberately align with value-sensitive system design methodologies[25] (Friedman et al., 2015). This standard, which is not specific to the transportation or technology industries, does not prescribe specific ethical tenets but rather focuses on an Ethical Values Elicitation and Prioritization process involving stakeholder groups. The standard notes within the definition of a stakeholder that "some stakeholders can have interests that oppose each other or oppose the system." Additionally, "groups of stakeholders: internal, users, opponents, and external authorities, are treated differently when risks, ethical values, and impacts are evaluated" (IEEE, 2021).

### 3.1 Exploring societal expectations around good driving behavior

Societal expectations can be explored through an established (and much debated) thought experiment. Philosopher John Rawls asked what rules a group of individuals would make if they did not know each of their own "place in society, class position, wealth, or social status, nor does anyone know their race, gender, fortune or misfortune in the distribution of natural assets and abilities, level of intelligence, strength, education, and the like."(Freeman, 2008). He argues that in such a situation, the principles that all individuals rationally agree to (or not reject) are fair, and that fairness extends to any institutions, laws, and conventions required by these principles. Consequently "the role of the social contract is to represent this idea, that the basic principles of social cooperation are justifiable hence acceptable [or not rejectable] to all reasonable and rational members of society, and that they are principles which all can commit themselves to support and comply with." (Freeman, 2008)

There is an interesting and real parallel in the transportation ecosystem, wherein we accept risks as a society but experience them as individuals (Tingvall & Lie, 2021). The average road user regularly plays multiple roles, moving from being a driver or a passenger in a car or bus, to a pedestrian (even if just walking from the transit stop, car park, or bicycle rack to their destination), to a cyclist, etc. often in the same week. Norms and expectations are generated by a collective, consensus understanding of many road users, each of whom uses the road in different ways at different times and who may shortly be occupying a different role.[26] Meeting said expectations becomes the realization of the social contract between road users. Consequently, societal norms and expectations, by their very nature, include the mutual agreement on acceptable driving behavior.

---

[24] What is considered acceptable and expected will differ depending on local driving culture. For example, what is considered an unexpected or aggressive "cut-off" during a lane change in one region, may be expected and necessary in another region.

[25] Value sensitive design is an approach to engineering design calling for the inclusion of broad stakeholders perspectives within the conceptualization of a system design, beyond those involved in the traditional development value-chain, to include those who will be impacted by the existence of a product (Himma & Tavani, 2008; Friedman et al., 1996)

Recent work in social philosophy distinguishes two types of expectations[27,28]:
- Empirical expectations: "beliefs that a certain behavior will be followed." (Bicchieri et al., 2023)

- Normative expectations: "beliefs that a certain behavior ought[29] to be followed." (Bicchieri et al., 2023)[30]

The source of these beliefs is not in scope for the present paper, but generally it is attributed that past experiences shape Empirical Expectations,[31] and agreed-upon principles shape Normative Expectations.[32,33] Additionally, Empirical Expectations are grounded in descriptive norms, while Normative Expectations are grounded in social norms (Bicchieri, 2006). As the reader may have gleaned from the definitions, these two sets are not disjoint. In fact, as presented below, Empirical Expectations can, over time, influence Normative ones. Additionally, Normative and Empirical Expectations can result in the same behavior (i.e., when what one has previously done is what one should do).

---

[26] For example, as we get out of a car, and walk a block to our destination, we transition from passenger to pedestrian.

[27] This is not to judge one type of expectation as being superior or preferable to another, but rather to draw a taxonomic set of distinctions between them, to aid discussion.

[28] Expectations are rooted in context; how we as a society form and seek the fulfillment of expectations around driving differs compared to flying or taking a train (Tingvall & Lie, 2021).

[29] "Ought" is used by Bicchieri to focus on expectations drawing on mutual respect; Bicchieri asks us to "consider… a rule of equal division. In this case we may believe that others ought to 'divide the cake in equal parts' because this is the fair thing to do. We think they have an obligation to follow the rule, a duty to be fair." (Bicchieri, 2006). Please note, a legal definition of duty is not being used by the author being quoted here.

[30] Normative Expectations have clear connections to the concepts set forth in IEEE 7000. For example, the standard uses the term personal maxims to describe a "personal principle of what one wishes for, acts upon, and thinks that it should be applicable to everyone." (IEEE, 2021)

[31] These expectations draw on past experiences, such that "one expects people to follow R in situations of type S because one has observed them to do just that over a long period of time. If the present situation is of type S, one can reasonably infer that, ceteris paribus, people will conform to R as they always did in the past." (Bicchieri, 2006)

[32] Normative expectations can be elicited from a wide range of sources, including the technical, peer-reviewed work and grey literature, rules of the road, regulations, training materials for new drivers, but also through structured interviews, casual conversations, conference presentations, etc. There is no one source for Normative Expectations.

[33] Empirical and Normative Expectations (confined by the laws of physics and technological capabilities) may vary based on stakeholder role and personal viewpoint (Pyrialakou et al., 2020; Othman, 2021; Yuen et al., 2020; Nair & Bhat, 2021).

Importantly, "the Normative expectations condition says that expectations are believed to be reciprocal. That is, not only do I expect others to conform, but I also believe they expect me to conform… In large, anonymous groups [as occurs on the roadway], if we do not want to act contrary to others' normative expectations it must be because we find such expectations reasonable" (Bicchieri, 2006). How the range of Normative Expectations on ADSs may differ from expectations for human drivers and human drivers supported by ADAS is an ongoing, industry-wide discussion (e.g., (Bonnefon et al., 2024; Hulse et al., 2018; De Freitas et al., 2021)). Additionally, Normative and Empirical Expectations on behaviors change, stemming from shifts both in the transportation ecosystem (e.g., the introduction of shared e-scooters) and beyond (e.g., changes in distracted driving rates due to cell phone usage).

Examples of discrepancies between these Expectations are rife in traffic safety. For example, it is an Empirical Expectation that some drivers are drunk (according to IIHS, 30% of all fatally injured drivers had a BAC> 0.08% in 2022 (IIHS, 2024a)), whereas it is a Normative Expectation that road users ought not to be drunk. Similarly, it is an Empirical Expectation that drivers speed (according to IIHS, there were 12,000 deaths in speed-related crashes in 2022 (IIHS, 2024b)); it is a Normative expectation to drive at a safe speed, anchored by the posted limit. Car seat usage for infants is an example of a behavior where the discrepancy between Empirical and Normative Expectation is now relatively small; according to a 2021 study, 98.8% of children under the age of 1 were in a car seat (Boyle, 2023). Minimizing discrepancies such that Empirical Expectations align with Normative Expectations supports improvements on the roadway. Furthermore, calibrating ADS behavior to support this minimization is the work of internal development and evaluation (a process outlined in Figure 2).

Empirical and Normative Expectations do not directly parallel the distinction between unsafe and safe behaviors. For example, well-intentioned stakeholders might have Normative Expectations, which if followed, could have unintended side-effects with negative safety repercussions (Fraade-Blanar et al, 2018). Consider, for instance, a situation in which hard brake rates dictate insurance premiums, inadvertently setting what could be perceived as a Normative Expectation that one should not hard brake. One could thus feel incentivized to limit hard brakes. However, research shows that hard braking is a common mechanism to avoid collisions; disincentivizing it may have unintended negative consequences.[34] Nuance, in general and specifically around possible unintended consequences, is required in considering the relationship between Normative Expectations and realized driving behavior. Additionally, it is possible for there to be multiple beliefs about a given behavior within a broad society, when there is no settled, dominant Normative Expectation (Bicchieri, 2006).[35, 36]

---

[34] As noted in Fraade-Blanar et al., (2018) recording of these events "would unfairly penalize the driver taking evasive maneuvers to avoid a crash" and could, at an extreme, disincentivize said avoidance. Guidance on how to proceed in such circumstances is limited; for example, the aforementioned ISO 39003 DR8-b notes the importance of acting in a manner that is 'optically' correct but, despite noting checks for internal consistency, fails to suggest how to proceed when this conflicts with what is safest (ISO, 2023).

[35] For example, a prevailing Normative Expectations has not been established around hands-free cell phone use while driving; specifically if talking on a handsfree device while driving is acceptable (the same survey found 19% believed it was dangerous and 42% supported a law prohibiting the behavior)(NCSL, 2024).

[36] Drivership specifically refers to Normative Expectations "that are followed by a majority of the target population, significantly affect behavior (intuitively, 'strong' norms), and do not fade over time without external intervention (intuitively, 'stable' norms)." (Bicchieri & Demo, 2024)

Beyond expectations of what will happen (based on past experience) and ought to happen (based on societal obligation), there is a third type of expectation:
- **Furtherance Expectations:** beliefs on which behavior(s) could be exhibited to enable continuous improvement of the transportation ecosystem, motivated by the intention to improve the status quo.

This type of Expectation stems from Drivership's contribution to roadway citizenship. It refers to expectations as opportunities to improve the road traffic system in terms of safety, efficiency, comfort, ease, etc. which could eventually be established as Normative Expectations in a society.[37]

All three types of expectations are constrained by what is possible and technologically/logistically/physically/etc. viable. It may be that for a given topic, Empirical, Normative, and Furtherance Expectations align or diverge (Figure 3). For example, if a low-severity collision occurs on a freeway, it may be an Empirical Expectation that the collision counterpart drives off (i.e. a hit-and-run) and the other party remains in place where the collision occurred. It may be a Normative Expectation that both parties pull off to the shoulder. However, researchers and law enforcement (e.g., Peiris & Harik, 2018; Texas Transportation Code, 2005; AZ DPS, 2022; ABC 30 Action News, 2021)) may consider that that it would cause less congestion and provide a safer option if both parties exited the freeway, if possible, and then pull over to reduce the chances secondary crash (either involving the original vehicle and another vehicle that comes upon the scene, or other vehicles who are maneuvering to avoid the scene). This behavior lives in the overlap between Furtherance and Normative Expectations; choosing to enact this last behavior represents the ongoing process of modifying expectations towards what is established as preferable based on mutual responsibility to all road users on the roadway, bearing in mind what is possible (e.g., the vehicles were drivable and there is an exit nearby).

Societal movements, regulatory efforts to shape/revise/enforce rules of the road, advocacy, etc. aim to pull Empirical Expectations to overlap with Normative Expectations by normalizing and then entrenching the desired behavior.[38] Careful, deliberate research and cross-functional discussion represent the process of pulling Normative Expectations into closer alignment with Furtherance Expectations. These three Expectations and what pulls them closer to or prevents them from aligning represents important topics for future research.

As society naturally shifts over time, expectations change. This includes changing conceptualizations on what will happen (Empirical), what should happen (Normative), and what could happen to improve the transportation ecosystem (Furtherance). Tensions may exist across the categories of Expectation (intuitively represented in Figure 3). We argue that Drivership operationalization is grounded in Normative Expectations that are shared at a societal level.[39] These Drivership Expectations have **Feasible Constraints**; regardless of their basis and/or source, these expectations are confined to what is currently technologically and physically (i.e., according to the laws of physics) achievable and practical.[40]

---

[37] What is involved in fulfilling an Expectation varies. This variation does not result in prioritization of expectations or denote quality or realism. Instead it speaks to what it means to achieve or exceed a given expectation and, by extension, an organization's role in said achievement. For example, for Furtherance Expectations like those inspired by Vision Zero, driving behavior is only a portion of the overarching concept which includes roadway design, vehicle design, etc. (Lie & Tingvall, 2024: Ecola et al., 2018). Vision Zero cannot be achieved through the efforts of any one company or organization as it entails shared responsibility. But a contribution to Vision Zero may be shown through, for example, a safety record, evidence of considering impact on the roadway as could be assessed by the FIA Foundation Road Safety Index, participating in safety-focused standards and best practices bodies, etc.

[38] For example, expectations around seat belt use (and subsequently, rates of seat belt use itself) changed as a result of the 70% by `92 program (CDC, 1992), the subsequent Click it or Ticket Campaign, etc. (Solomon et al., 2004; Nichols & Solomon, 2013).

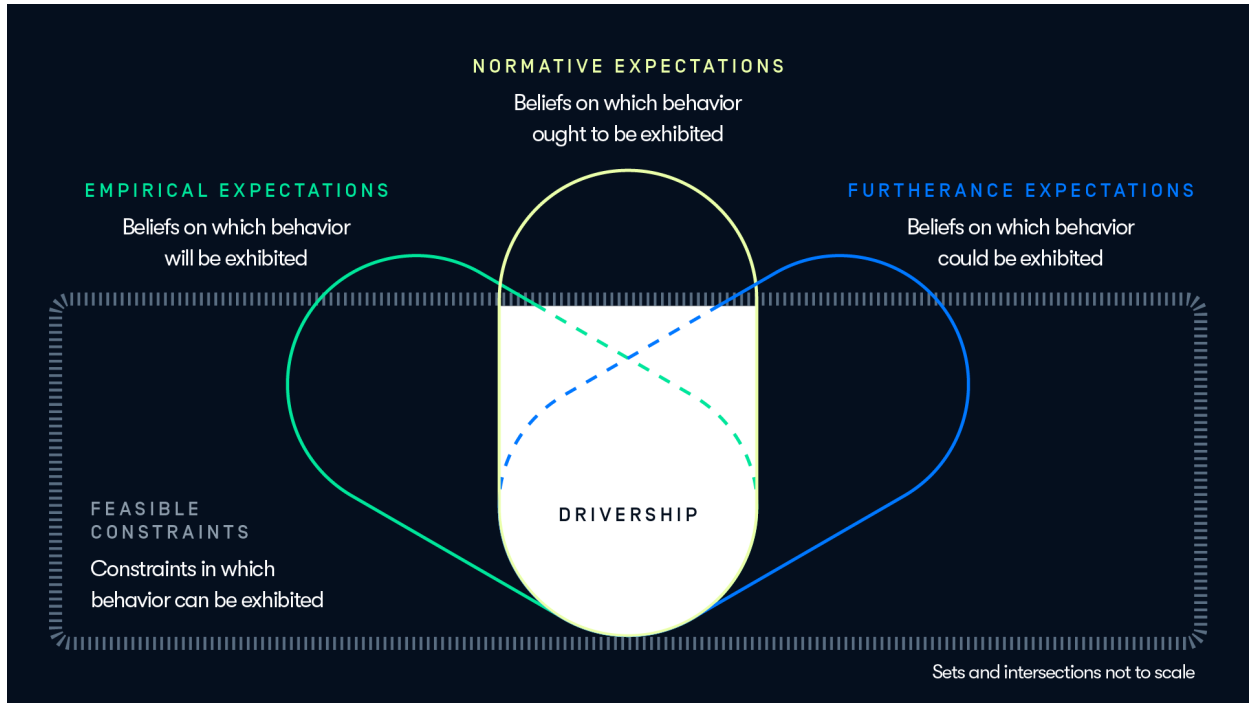

Figure 3. Drivership (in purple) in relationship to societal Normative, Empirical, and Furtherance Expectations

A sound conceptualization of Drivership thus needs to ensure that the attributes through which it is evaluated reflect current, societal, feasible Normative Expectations. Operationalization occurs with an understanding that non-Drivership expectations may exist (i.e., non-Normative Expectations, Normative Expectations that are beyond the Feasibility Constraints, Expectations that do not exist at the societal levels, etc.) and that the ADS is only one of many road users.

---

[39] This situation of Drivership Expectations at the societal level clearly recalls its roots in Contractualism; focusing a "set of principles for the general regulation of behaviour that no one could reasonably reject as a basis for informed, unforced, general agreement," (Scanlon, 1998) and motivated by a combination of "self-regard and by respect for others." (Ashford & Mulgan, 2007)

[40] An excellent illustration of Feasible Constraints can be found in stopping distances. In 2021, according to Consumer Reports, the average stopping distance (from 60mph) for passenger vehicles was 132 feet (Bartlett, 2021). A more recent article noted the Ford Mustang Dark Horse set a new record with a stopping distance (from 60mph) of 86 feet (which involved an average declaration of 1.4Gs) (Dushyant, 2023). Consequently, an expectation of a distance of 75 feet between an obstacle being introduced and the vehicle coming to a full stop from 60mph would be outside of the Feasible Constraints, because it is not technologically possible currently. An expectation of 125 feet would be inside the Feasibility Constraint, while a stopping distance of 175 feet would not be Normative.

**3.2 Towards the operationalization of Drivership**
The operationalization of Drivership[41] may require a range of different methodologies, including leveraging expert judgment, in combination, where available, with quantitative behavior reference models (like those listed in section 2.1). These models have benchmarks calibrated to or based on driving qualities that meet societal expectations.

One example of such a behavior reference model, applicable to the expectation of responding properly to potential and actual conflicts, is the *NIEON model.* This model is intended to represent the collision avoidance response performance expected from a Non-impaired human driver with their Eyes ON the conflict. Engström et al. (2024) describes how such a quantitative response model can be fitted to human data from naturalistic driving that fulfill the NIEON criterion and Scanlon et al., (2022) demonstrate how it can be used as a benchmark in simulated collision avoidance testing of ADSs. The NIEON model is thus intended to represent Normative Expectations on collision avoidance performance, as a benchmark for ADS performance. This can be contrasted to a collision avoidance driver model designed to represent Empirical Expectations, that is, how drivers in the general population (some of which may be distracted, impaired, etc.) are expected to respond to conflicts. Such a model may, for example, provide a probabilistic representation of the collision avoidance performance of the general driver population (for example, Bärgman et al., 2015; Engström et al., 2023).

Another example of a reference model is Surprise, described in Dinparastdjadid et al., (2023). This is a novel type of model applicable to evaluating the expectation of avoiding initiating conflicts. More specifically, the Surprise model can be used to evaluate the degree of surprise associated with different maneuvers such as braking and lane changing, as experienced by other road users. The model is based on comparing the predicted behavior of the ego vehicle to the actual behavior (i.e., probabilistic mismatch surprise) or, alternatively, how much a current belief about the future behavior of the ego vehicle changes with new observations (i.e., belief mismatch surprise). Since traffic conflicts are typically initiated by behaviors that surprise other road users, avoiding surprising behaviors is key to avoid initiating conflicts.

Other types of benchmarking could apply to rules of the road's compliance expectations. The evaluation of compliance with rules of the road remains among the fundamental tenets of Drivership; within the framing of Figure 3, one could explore setting internal benchmarks to evaluate the compliance-focused appropriateness of automated vehicle behaviors. On one hand, Empirical Expectations grounded in the observation of pervasive violations on the part of human drivers would likely offer low rigor for establishing an acceptable benchmark.[42] On the other hand, Furtherance Expectations pushing for a perfect record based on hard coding of individual rules may discount emergency situations in which rules are violated to preserve safety or in response to instructions from officers on the road.[43] A composite benchmark could thus be devised to combine the verification of ability and proficiency (for example, through scenario-based testing) with appropriate product governance and management processes based on internal monitoring of on-road operations.[44] These processes could be informed by the setting of feasible (recalling feasibility constraints) and acceptably low rates of occurrence of pre-specified events of interest. Reference models could then support the devising of behavioral criteria for compliance capabilities: for example, evaluating timing decisions to proceed at a yellow light, or stopping location in relation to the geo-specific normative stop line.

---

[41] The evaluation of Drivership is highly context dependent and involves a holistic understanding of the spectrum of behaviors that each traffic participant could have undertaken and the entirety of the scenario timeline. In evaluating, it may prove helpful to speak of Micro Drivership (i.e., analysis of a specific road user maneuver) and Macro Drivership (i.e., analysis of the chains of maneuvers by different road users, bearing in mind the broader goals and all the alternatives available to the different road users, long-term implications, counterfactuals, etc.). It may be that behavior within a scenario, when taken more broadly, contributes to a larger goal of safety specifically and goodness broadly, but one specific maneuver appears to be poor (Micro) Drivership. Essentially, "the wrongness of an action depends on the circumstances in which it is performed." (Scanlon, 2010)

[42] Of note, in actual practice, there are many reasons human road users may make an intentional road rule violation (which is distinct from an (unintentional) mistake or slip (Parker et al., 2007; Reason, 1990)). As an extreme example, a survey of bicyclists in Montreal noted that only 0.6% chose behavior that consistently followed traffic laws, with ''it was the safest option,'' or ''it made sure I was visible to cars'' as the most common reason for not following traffic laws (Chaloux & Ahmed, 2019).

Figure 4 visualizes the role that behavior reference models can play within the evaluation of Drivership, and the connection to societal expectations. This figure shows how the socially aware and safety-centric qualities of driving behavior are assessed, by individual or overlapping behavior reference models. On the right is a taxonomic disaggregation of Drivership, the discrete components of which can, in turn, be evaluated through the use of behavior reference models (at the bottom of the visual). On the left is the flow illustrating how Empirical, Normative, and Furtherance Expectations reciprocally relate to this decomposition. Of note, the operationalization of Drivership - as grounded in feasible Normative Expectations - involves the precise definition of behavioral requirements for the ADS derived from such Expectations. Conversely, the specification of behavioral reference models stems from such behavioral requirements. Two-way arrows in the figure denote the mutualistic nature, as previously presented in Figure 2.

---

[43] See, for example, the CA DMV provision of Article 3.7 §227.32 (c), stating "except when necessary for the safety of vehicle's occupants and/or other road users." (CA DMV, 2022)

[44] This is aligned with the approach Waymo presented in its 2020 paper overviewing the role of various methodologies for safety readiness determination (Webb et al., 2020).

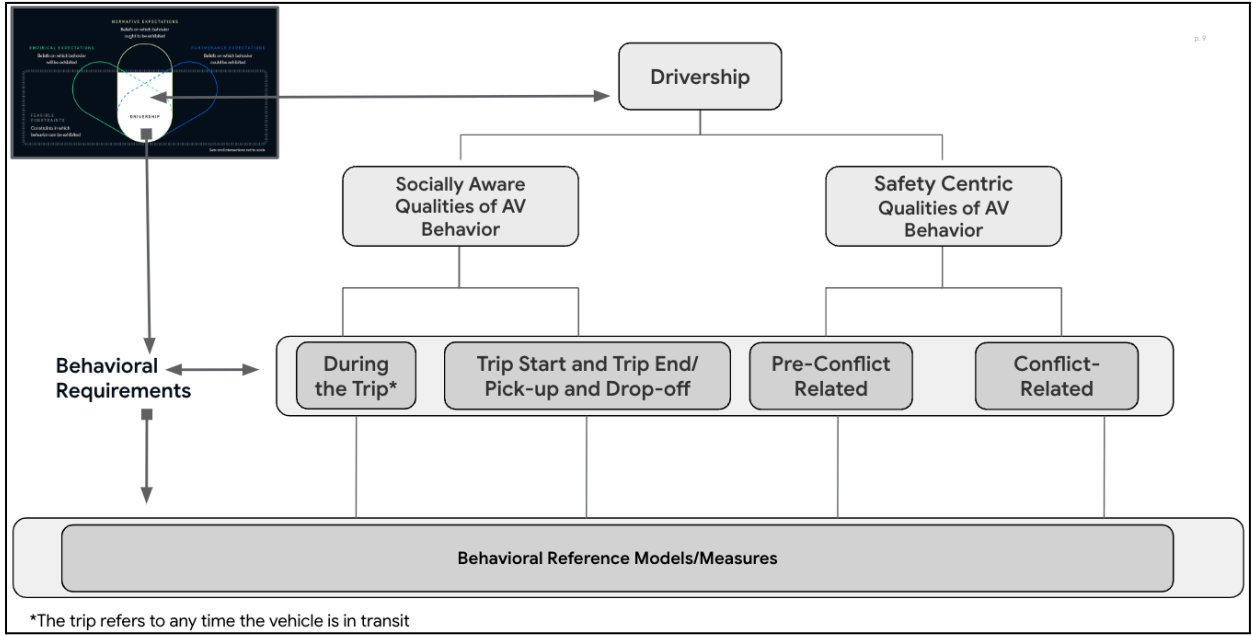

Figure 4. A broad framework of the operationalization of Drivership, in connection with Expectations

## 4. What is the Relationship between Drivership and a Safety Case?

Having established that Drivership embeds both considerations of safety and social awareness, the question then becomes what is the connection between Drivership and a safety case. A safety case is a "structured argument, supported by a body of evidence that provides a compelling, comprehensible, and valid case that a system is safe for a given application in a given environment" (UL 4600, 2023). Based on the standardized definition of "safety" (ISO, 2018; ISO, 2022; Favarò et al., 2023), the determination of Absence of Unreasonable Risk (AUR)[45] is established as the top-level goal. Determining Absence of Unreasonable Risk involves assessing risk from behavioral, operational, and architectural hazards[46] (Favarò et al., 2023). For behavioral hazards specifically, Drivership provides the means to assess reasonableness of behavioral risk in connection to valid societal moral concepts (ISO, 2018, 2022), denoted here as feasible Normative Expectations. Consequently, Drivership supports identification of the types of driving events most relevant for the behavioral component of Absence of Unreasonable Risk (AUR). Figure 5 visualizes the intersection between improper Drivership and risk of harmful outcomes in interactions between road users.

---

[45] An unreasonable risk is a "risk judged to be unacceptable in a certain context according to valid societal moral concepts." (ISO, 2018)

[46] Per ISO 26262, a hazard is a "potential source of harm" (ISO, 2018). Within the context of Drivership, we focus on behavioral hazards as those associated with potential sources of harm resulting from the ADS's displayed driving behavior. (Favaro et al., 2023)

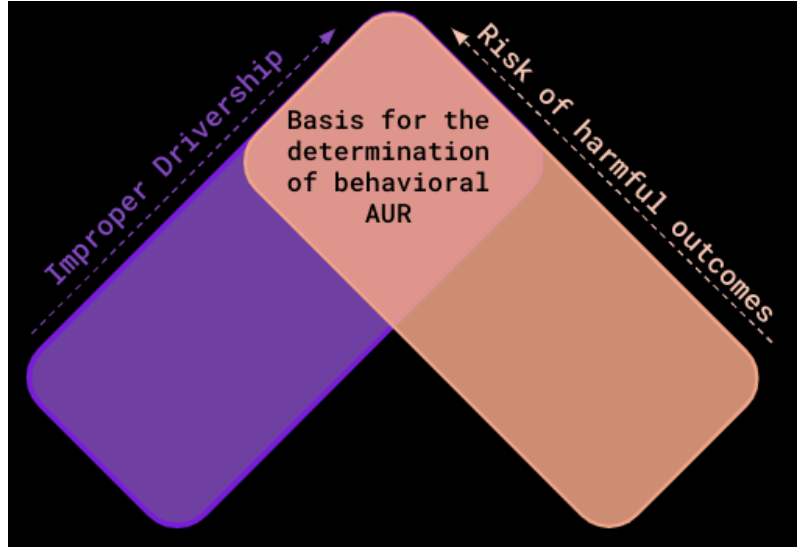

Figure 5. The intersection of improper Drivership and the risk of harmful outcomes

While behavior involving improper Drivership (represented in lavender in Figure 5) and behavior where there is a risk of harmful outcomes (represented in grey in Figure 5) should be continuously analyzed as part of engineering improvement practices, they should be appropriately differentiated. Specific to behavioral AUR, it is important to differentiate interactions where (A) there is improper Drivership but no risk of harmful outcomes, from (B) the behavior does not represent improper Drivership but, as a result of the larger interaction in which it occurs, it carries a safety risk, from (C) there is improper Drivership involving a risk of harmful outcomes.

Drivership's inclusion of safety-centric and socially-aware behavior can add specificity to safety evaluation practices along two directions:
1. Risk can still be regarded as *reasonable* if the driver exhibits proper (even if non-ideal) Drivership, and even if a high severity collision results. Evaluation provides a needed lens to identify events wherein there is both a risk of harmful outcomes and improper Drivership; this lens allows appropriate prioritization safety-critical events, differentiating, for example, the relevance of conflicts initiated by other road users which still showcase proper Drivership by the responder (e.g., their proper response resulted in the successful mitigation of outcomes to a lower degree of severity (as illustrated by (B) in Figure 5)).
2. Improper Drivership, which results from a misalignment between driving behavior and Normative Expectations (as illustrated in Figure 2 and 3), may not, on its own, be a sufficient tell of a safety concern. This is nested in the very definition of socially-aware attributes of Drivership (as described in Section *3. What is Drivership?*), which are more distal to harm manifestation and without a clear or proximal injury risk.

Behavior reference models and methodologies used to evaluate Drivership contribute to the evaluation of Absence of Unreasonable Risk within a safety case, by assessing the relationship between on-road behavioral hazards and societal expectations. Thus, evaluating Drivership is key to assessing the extent to which the risk associated with a given behavioral hazard is unreasonable[47] (Favarò et al., 2023). For Drivership, the evaluation of a specific event would go beyond the contribution to a safety determination to consider appropriateness of and societal expectations around the behavior.

---

[47] Please note, a legal definition of reasonableness or unreasonableness is not being used here.

Finally, because Drivership includes but is not limited to safety, it is considered in all stages of the Safety Determination Lifecycle (Favarò et al., 2023; AVSC, 2024). Herein, Drivership serves to inform product behavioral development cycles, as well as the appropriate setup of coherent internal processes and methodologies to aid credibility and confidence into the performance evaluation as it may contribute to a safety case.

## 5. Conclusion: how to move forward

A key question motivates Drivership: what is good driving behavior? This question concerns not just ADSs but rather all involved in the field of motorized transportation, which has been enmeshed in society for over a century. It is intuitively clear that there is no objective answer to this question. Our survey of the subjective answers of other researchers and policymakers unearthed a wide variety of topics, values, principles, and models, without either a clear path to operationalization nor an overarching framing. Additionally, these answers frequently lacked a clear justification for the inclusion or exclusion of certain attributes or perspectives.

We thus began by refining the question on good driving behavior to: what do road users owe each other (Scanlon, 1998)? This iteration, in turn, led us to analyze stakeholders expectations, where, based upon prior literature, we taxonomically distinguished between: Empirical, Normative, and Furtherance Expectations. We chose to anchor Drivership in Normative Expectations at the societal level, limited by what is feasible; **in the simplest of terms, good driving behavior as the behavior that a driver ought to do and plausibly can do, to be a good citizen of the roadway (feasible Normative Expectations).** This may or may not overlap with the behavior they are anticipated to do (Empirical Expectations), and/or could do in order to improve the transportation ecosystem (Furtherance Expectations).

We offer a path toward operationalization by discussing how, among other tools, Normative Expectations can be used as the basis for the development of benchmarks implemented by behavior reference models. We also sought to distinguish qualities of goodness in driving behavior; safe, of course, but also socially aware. This distinction is important when connecting Drivership to a safety case, which requires in-depth evaluation of those events that combine both improper Drivership and potential for injury risk

**Drivership is a realization of good driving behavior, providing a pathway to achieving this goal.** Growing and maintaining a Drivership approach includes:
1. Developing and supporting the architecture and methodologies that contribute to a determination of Drivership;
2. Generating and supporting an evolving understanding of dynamic Normative Expectations with the public, policy-makers, and other key stakeholders to support operationalization;
3. Analyzing events to understand alignment between behavior and societal expectations;
4. Planning for continuous Drivership improvement to be crafted, according to realistic expectations on engineering capabilities and technical challenges and constraints;
5. Across stakeholders (and beyond developers), engaging in consensus-creating-and-confirming activities regarding Normative Expectations on ADS behavior.

This represents a cyclic process that supports a value-based system approach. Understanding Empirical, Normative, and Furtherance Expectations represents not a one-time occurrence but an ongoing, progressive process of change toward the world we want to live in; evolving answers to the question of good driving behavior requires regular re-alignment into the future. Further research is needed to develop Drivership into a technology-, and autonomy-neutral approach to understanding good driving behavior. Advancing a mutualistic Drivership framing provides definition and supports evaluating "goodness" of behavior, from the road users involved in Mr. Bliss's death in New York 125 years ago to those seen today, and into tomorrow.